\documentclass[twocolumn]{aastex631}

\usepackage[T1]{fontenc}
\usepackage{hyperref}
\usepackage{amsmath}

\defcitealias{schlaufman2018evidence}{S18}

\accepted{to the Astronomical Journal}

\shortauthors{Giacalone et al.}

\graphicspath{{./}{figures/}}
\begin{document}

\title{The Transition from Giant Planets to Brown Dwarfs beyond 1 au from the Stellar Metallicity Distribution}

\correspondingauthor{Steven Giacalone}
\email{giacalone@astro.caltech.edu}

\author[0000-0002-8965-3969]{Steven Giacalone}
\altaffiliation{NSF Astronomy and Astrophysics Postdoctoral Fellow}
\affiliation{Department of Astronomy, California Institute of Technology, Pasadena, CA 91125, USA}

\author[0000-0001-8638-0320]{Andrew W. Howard}
\affiliation{Department of Astronomy, California Institute of Technology, Pasadena, CA 91125, USA}

\author[0000-0003-0742-1660]{Gregory J. Gilbert}
\affiliation{Department of Physics \& Astronomy, University of California Los Angeles, Los Angeles, CA 90095, USA}

\author[0000-0002-4290-6826]{Judah Van Zandt}
\altaffiliation{NASA FINESST Fellow}
\affiliation{Department of Physics \& Astronomy, University of California Los Angeles, Los Angeles, CA 90095, USA}

\author[0000-0003-0967-2893]{Erik A. Petigura}
\affiliation{Department of Physics \& Astronomy, University of California Los Angeles, Los Angeles, CA 90095, USA}

\author[0000-0002-9305-5101]{Luke B. Handley}
\affiliation{Department of Astronomy, California Institute of Technology, Pasadena, CA 91125, USA}

% \author[0000-0002-0531-1073]{Howard Isaacson}
% \affiliation{501 Campbell Hall, University of California at Berkeley, Berkeley, CA 94720, USA}
% \affiliation{Centre for Astrophysics, University of Southern Queensland, Toowoomba, QLD, Australia}

%\author[0000-0003-0967-2893]{Erik A. Petigura}
%\affiliation{Department of Physics \& Astronomy, University of California Los Angeles, Los Angeles, CA 90095, USA}



\begin{abstract}

Giant planets and brown dwarfs are thought to form via a combination of pathways, including bottom-up mechanisms in which gas is accreted onto a solid core and top-down mechanisms in which gas collapses directly into a gravitationally-bound object. One can distinguish the prevalence of these mechanisms using host star metallicities. Bottom-up formation thrives in metal-rich environments, whereas top-down formation is weakly dependent on ambient metal content. Using a hierarchical Bayesian model and the results of the California Legacy Survey (CLS), a low-bias and homogeneously analyzed radial velocity survey, we find evidence for a transition in the stellar metallicity distribution at a companion mass of $\gamma = 27_{-8}^{+12} \, M_{\rm Jup}$ for companions with orbital separations between $1-50$~au. Companions below and above this threshold tend to orbit stars with higher ($\rm{[Fe/H]} = 0.17 \pm 0.12$ dex) and lower ($\rm{[Fe/H]} = -0.03 \pm 0.10$ dex) metallicities, respectively. %In addition, we find that the two populations have distinct orbital eccentricity distributions, with the low-mass population (the ``planets'') more strongly preferring low-eccentricity orbits. 
{ Previous studies of relatively close-in companions reported evidence of a lower transition mass of $\leq 10 \, {\rm M_{\rm Jup}}$. When applied to the CLS sample, our model predicts the probability of a transition in the stellar metallicity distribution at or below $10 \, { M_{\rm Jup}}$ to be $< 1 \%$. We compare our results to estimates of $\gamma$ gleaned from other observational metrics and discuss implications for planet formation theory.}

\end{abstract}



\section{Introduction} \label{sec:intro}

In the past, the classification of an object as a giant planet or a brown dwarf was solely dependent on its mass in relation to the deuterium-burning limit ($\sim 13 \, M_{\rm Jup}$; \citealt{saumon1996browndwarf, chabrier2000browndwarf, spiegel2011deuterium, bodenheimer2013deuterium}): those with masses below the limit were considered giant planets and those with masses above the limit were considered brown dwarfs \citep{boss2007deuterium}. However, it has been argued that the classifications of substellar companions should instead depend on formation mechanism, where only objects that form ``like planets'' are considered planets \citep[e.g.,][]{burrows2001browndwarf, luhman2008browndwarf, chabrier2014formation}. Objects that form like planets are those that form bottom-up, beginning with the aggregation of a solid core onto which gas is accreted from the surrounding protoplanetary disk. This core-accretion scenario has long stood as the most likely formation mechanism for gas giant planets and is believed to be able to form objects with masses several times that of Jupiter \citep{pollack1996coreaccretion, alibert2005core, hubickyi2005coreaccretion, matsuo2007coreaccretion}. Conversely, objects that form top-down, through the direct collapse of gas without a preexisting solid core, are considered to form ``like stars'' \citep{bate2012formation, kratter2016gi, offner2023origin}. Despite this definition, top-down formation mechanisms are thought to be capable of forming companions with masses comparable to that of Jupiter in certain environments \citep[e.g.,][]{diamond2024pmo, fu2024pmo}. The top-down formation mechanism of disk fragmentation (also referred to as ``gravitational instability''), for instance, can form substellar companions with Jupiter-like and greater masses at wide orbital separations \citep{adams1992instability, boss1997GI, durisen2007gi, kratter2010gi, zhu2012GI}. These predictions make it challenging to determine how substellar companions form based on their masses alone.

Stellar metallicity, which correlates with the metallicity of the protoplanetary disk and is known to influence planet demographics, can likely help disentangle these formation mechanisms \citep[e.g.,][]{udry2007, buchhave2012metallicity, buchhave2014metallicity, winn2017metallicity, petigura2018metallicity, boley2021halo, boley2024metallicity}. Formation via core accretion is predicted to be more efficient in metal-rich environments \citep{ida2004model}. This has been supported by early radial velocity surveys, which have found giant planets more likely to exist around stars with higher metallicities \citep{santos2004metallicity, fischer2005metallicity, sousa2011}. Top-down formation is not believed to share this quality, with most mechanisms being indifferent to ambient metallicity \citep{boss2002metallicity}. Thus, by examining the distribution of substellar companions in the plane of mass and stellar metallicity, we can begin to discern the pathways responsible for their origins.

\begin{figure*}[t!]
  \centering
    \includegraphics[width=0.98\textwidth]{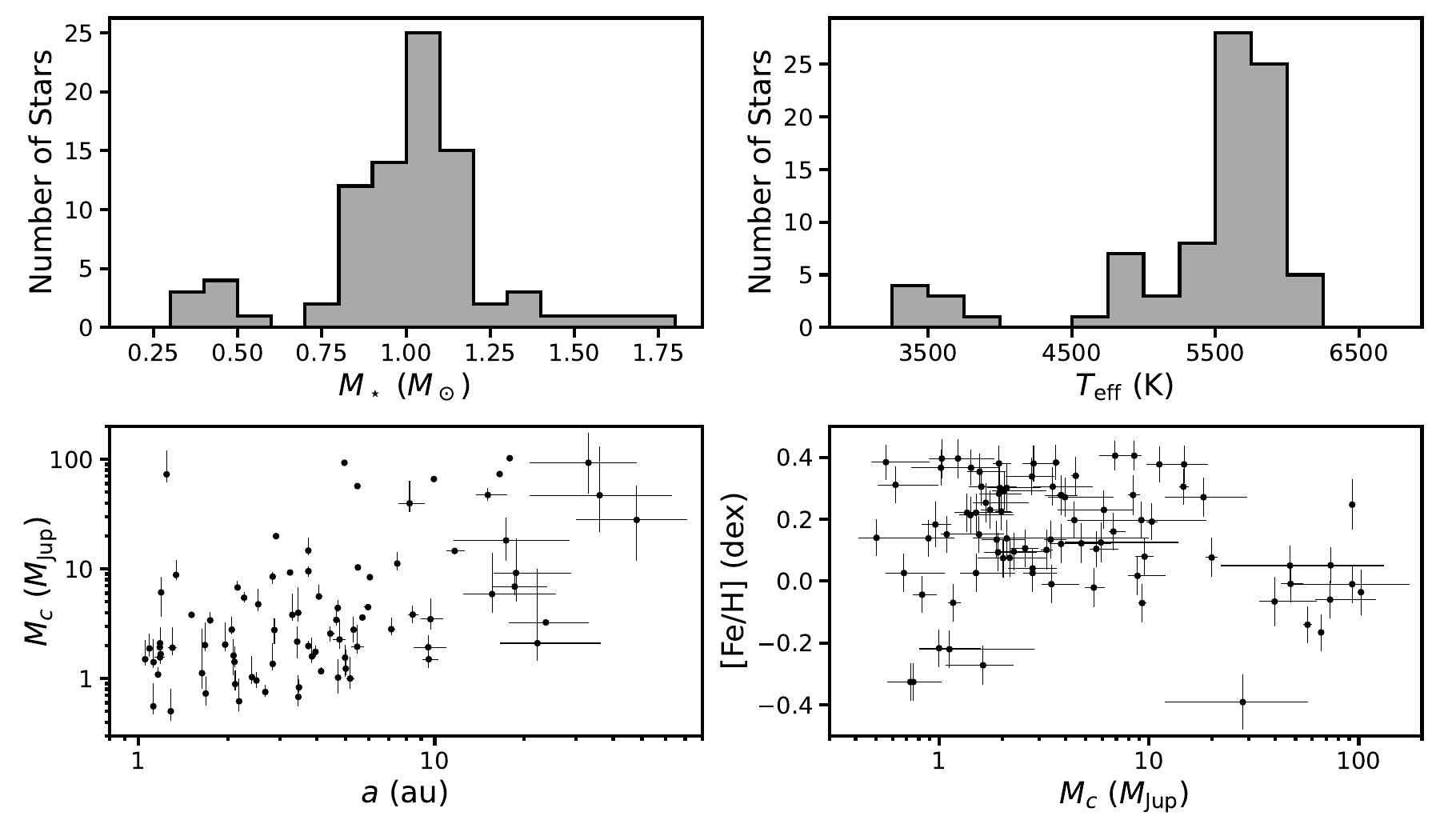}
    \caption{Visualization of the sample of CLS systems analyzed in this paper. The top two panels are histograms of the masses (top left) and effective temperatures (top right) of the stars in the sample. The bottom two panels show the distribution of systems in the orbital separation -- companion mass plane (bottom left) and the companion mass -- stellar metallicity plane (bottom right).}
    \label{fig: 1}
\end{figure*}

A number of studies have explored the differences in the metallicity distributions of giant planet and brown dwarf host stars using a variety of methods and observational samples. \citet{ma2014statistical} used a Kolmogorov-Smirnov test to show that companions with masses between $1 \, M_{\rm Jup}$ and $5 \, M_{\rm Jup}$ orbit stars with a distinct metallicity distribution from those with masses above $13 \, M_{\rm Jup}$. Evidence for the presence of two distinct host star metallicity distributions were also found by \citet{matasanchez2014abundances} and \citet{maldonado2017chemical}. \citet{santos2017evidence} used a multivariate Gaussian clustering technique on a sample of companions primarily detected by radial velocity surveys and found evidence for a transition in the stellar metallicity distribution around a minimum mass ($M_c{\rm sin}(i)$, where $M_c$ is the mass of the companion and $i$ is the inclination of the orbit) of $4 \, M_{\rm Jup}$, where less massive companions tend or orbit stars that are more metal rich. This analysis was performed using homogeneously determined stellar metallicities from the SWEET-Cat database \citep{santos2013sweetcat} and companions with orbital periods mostly under 1~yr (i.e., orbital periods under 365~days for $88.75 \%$ of the sample). \citet[][hereafter S18]{schlaufman2018evidence} performed a similar analysis using transiting giant planets, brown dwarfs, and low-mass stars with close-in orbits (i.e., semi-major axes $a < 0.25$~au for $95\%$ of the sample) and found that the transition between populations most likely occurs between $4$ and $10 \, M_{\rm Jup}$. { Most recently, \citet{wang2025accretion} investigated the metallicities of wide-separation ($a > 10$~au), directly imaged giant planets and brown dwarfs and found that companions with masses under $10 \, M_{\rm Jup}$ tend to have super-stellar metallicities and companions with masses over $20 \, M_{\rm Jup}$ tend to have metallicies similar to their host stars, suggesting a transition between $10$ and $20 \, M_{\rm Jup}$.}

In the present paper, we revisit the question of whether formation mechanisms for giant planets and brown dwarfs can be distinguished via the companion mass -- stellar metallicity distribution. Our analysis differs from previous studies in a few ways. First, we use a sample of uniformly characterized systems from a single, robustly assembled observational survey. This approach allows us to avoid biases introduced by compiling a sample of objects from different surveys, which may have employed different target selection functions and analysis techniques. Second, our sample consists only of objects with { $a = 1 - 50$~au, probing a distinct range of orbital separations compared to} previous works. Lastly, we { utilize a hierarchical} Bayesian approach for modeling the transition between two distinct stellar metallicity distributions as a function of companion mass. Previous techniques, by comparison, relied on transiting { or directly imaged} planets to estimate the transition mass and have only been capable of estimating a singular value or a simple range of values. For complementary analyses of the transition between giant planets and brown dwarfs based on orbital eccentricities and occurrence rates, we direct the reader to { \citet{gilbert2025eccentricities} and \citet{vanzandt2025cls}}.

This paper is structured as follows. In Section \ref{sec:sample}, we define the sample that we analyze. In Section \ref{sec:analysis}, we outline our analysis approach and report a new estimate of the giant planet--brown dwarf transition mass based on stellar metallicities. In Section \ref{sec:discussion}, we compare our estimate of transition mass with those made using other techniques and observational signatures. Lastly, in Section \ref{sec:conclusion}, we provide concluding remarks.

\section{Sample} \label{sec:sample}

We compiled our sample of companions from the California Legacy Survey (CLS; \citealt{rosenthal2021CLS, rosenthal2022CLS, rosenthal2024CLS, fulton2021CLS, isaacson2024CLS}). This decades-long survey collected radial velocities of 719 nearby FGKM-type stars with the HIRES spectrograph on the 10-meter Keck-I Telescope, detecting 226 objects with $M_c{\rm sin}(i)$ between 0.01 and 1000~$M_{\rm J}$ \citep{rosenthal2021CLS}. In addition, this survey performed a homogeneous characterization of all stars in the sample using \texttt{SpecMatch} \citep{petigura2015PhDT, yee2017specmatchemp} and \texttt{Isoclassify} \citep{huber2017isoclassify, berger2020isoclassify, berger2023isoclassify}, reporting stellar mass ($M_\star$), stellar radius ($R_\star$), stellar surface gravity ($\log g$), stellar effective temperature ($T_{\rm eff}$), and stellar metallicity ([Fe/H]) for each target. Because the survey was not biased towards stars more likely to host planets and analyzed its data set in a homogeneous manner, the CLS sample provides a powerful opportunity to robustly investigate planet demographics \citep[e.g.,][]{zink2023hj, vanzandt2024enhancement}.

{ Recently, \citet{vanzandt2025cls} generated posterior distributions for the true masses of 192 of the companions detected in CLS (all companions with $M_c {\rm sin}(i) < 80 \, M_{\rm Jup}$) by performing joint orbital fits on the radial velocity data and {\it Gaia/Hipparcos} astrometry. We used these $M_c$ posterior distributions in our analysis, such that we can avoid needing to marginalize over orbital inclination.}

We selected all systems that have companions with $0.5 \leq  M_c /M_{\rm Jup} \leq 110$ and with $a \geq 1$~au. Our imposed lower limit on $M_c$ was selected to avoid including companions that were unlikely to have formed via core accretion. In addition, by limiting our sample to very massive objects, we avoid potential biases caused by survey incompleteness; the CLS dataset is nearly $100\%$ sensitive to companions in this mass range (see \citealt{rosenthal2021CLS} and \citealt{vanzandt2025cls}).  Our imposed limits on $a$ were chosen because the CLS found very few objects with $M_c > 10 \, M_{\rm Jup}$ within 1~au. The limits therefore restricted the sample to orbital separations in which companions spanning nearly the full mass range were detected. { Lastly, following \citet{vanzandt2025cls}, we excluded HD 167215~c from the analysis, due to the companion being a likely false positive. In addition, we omit HIP 57050~d and HD 45184~d due to these companions having highly unconstrained masses.} Our sample of 85 companions is shown in Figure \ref{fig: 1}. We note that all of the stars in the sample have $T_{\rm eff} < 6500$~K and $80\%$ having $T_{\rm eff}$ between 5000 and 6200~K. This is an artifact of the selection function used by CLS, which avoided hot stars due to rapid rotations that limit radial velocity precision.

\begin{figure}[t!]
  \centering
  \hspace{-0.6cm}
    \includegraphics[width=0.5\textwidth]{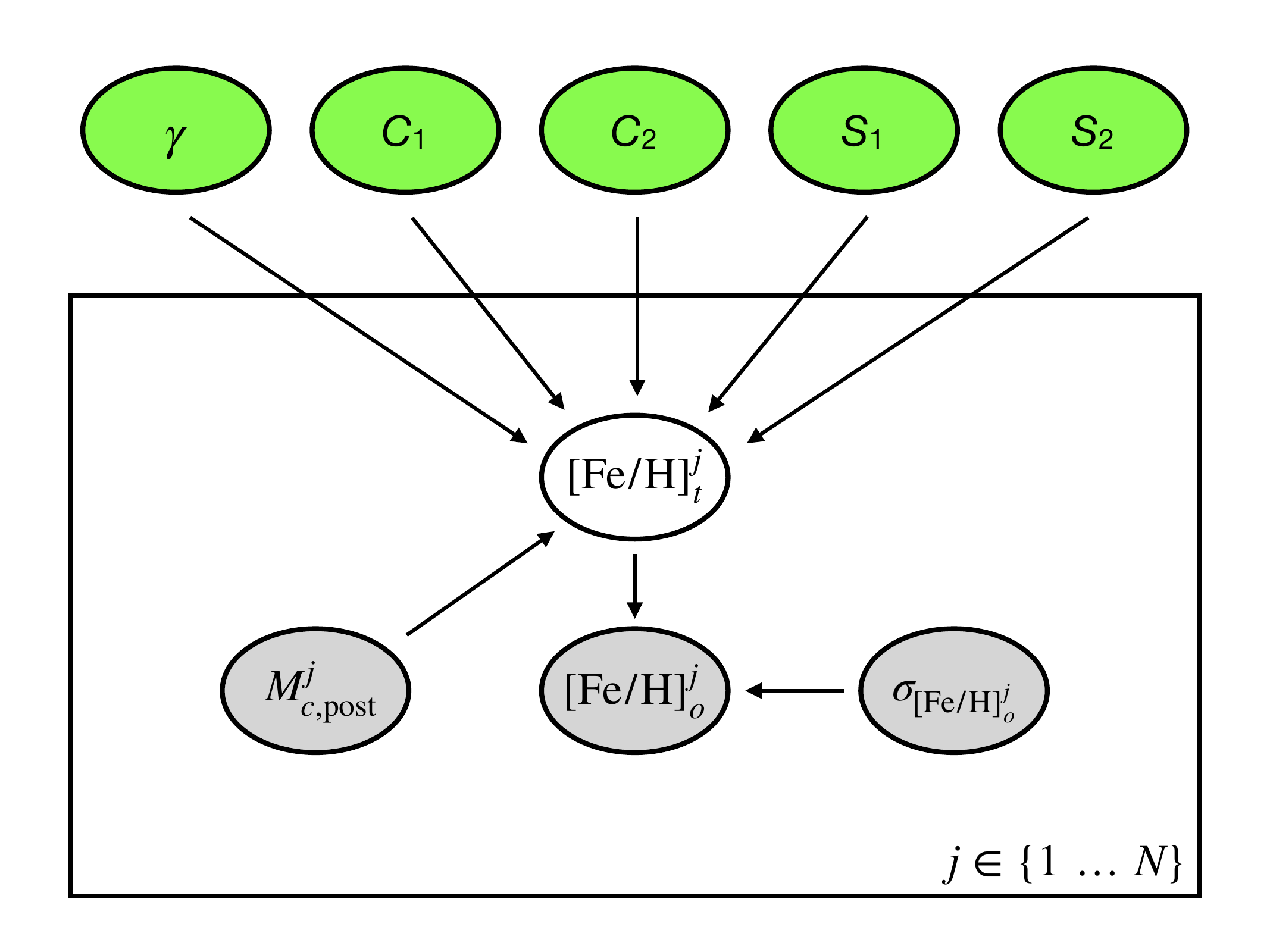}
    \caption{ Plate diagram illustrating our HBM framework. Latent parameters (white) and observed parameters (grey) are located inside the box. The latent and observed parameters, indexed via the superscript $j$, have $N$ members corresponding to the number of systems in the sample. Hyperparameters (green) are located outside the box. The model parameters are as follows: companion mass between which the stellar metallicity distributions transition ($\gamma$), low-mass population metallicity distribution mean ($C_1$), low-mass population metallicity distribution standard deviation ($S_1$), high-mass population metallicity distribution mean ($C_2$), high-mass population metallicity distribution standard deviation ($S_2$), companion mass posterior distribution ($M_{c, {\rm post}}^j$), observed stellar metallicity (${\rm [Fe/H]}_o^j$), and the uncertainty of the observed stellar metallicity ($\sigma_{{\rm [Fe/H]}_o^j}$).}
    \label{fig: plate}
\end{figure}

\section{Analysis} \label{sec:analysis}

In this section, we outline our analysis of the sample in the $M_c - {\rm [Fe/H]}$ plane. Upon visual inspection of the bottom right panel in Figure~\ref{fig: 1}, we noticed that the distribution of [Fe/H] appeared to change beyond $M_c \sim 20 \, M_{\rm Jup}$, but it was is not immediately clear if this change was statistically significant. As we detail below, we characterized this apparent change using a { hierarchical} Bayesian framework and tested its statistical significance.

\subsection{ Hierarchical Bayesian Model}\label{sec: modelfit}

% to do: update notation and write likelihood function more clearly (remove current eq 1 and 4)
% replace "true mass" with just "mass"

{ We utilize a hierarchical Bayesian model (HBM) to explore the transition in [Fe/H] distributions \citep[e.g.,][]{rogers2015most, wolfgang2016prob, neil2020bayesian, dong2023bayesian, siegel2023perp, schulze2024gap}. HBMs are particularly well suited for making population-level inferences from limited data sets, especially in cases where there are population-level ``hyperparameters'' that are dependent on both observed parameters and unobserved parameters for which there are well-known prior distributions (hereafter ``latent parameters'').\footnote{Note that not all population-level parameters are hyperparameters, but in this case the hyperparameters are the population-level parameters that we care about.} In our case, the observed parameters are the posterior distributions of the companion mass ($M_{c, {\rm post}}$), the observed metallicity (${\rm [Fe/H]}_o$), and the uncertainty in the observed metallicity ($\sigma_{{\rm [Fe/H]}_o}$). The latent parameter in our model is the true stellar metallicity (${\rm [Fe/H]}_t$). Lastly, the hyperparameters are the true mass at which the true [Fe/H] distribution changes ($\gamma$), the mean and standard deviation of the true [Fe/H] distribution for the low-mass population ($C_1$ and $S_1$, respectively), and the mean and standard deviation of the true [Fe/H] distribution for the high-mass population ($C_2$ and $S_2$, respectively). The HBM framework is displayed in Figure~\ref{fig: plate}.}

\begin{deluxetable*}{lllc}[t!]\label{tab: results}
% \tabletypesize{\footnotesize}
\tablewidth{\textwidth}
 \tablecaption{Hyperparameters of our HBM, their priors, and their posteriors.}
 \tablehead{ 
 \colhead{Description} & \colhead{Parameter} &  \colhead{Prior} & \colhead{Posterior}
 }
\startdata
Transition mass (in units of $M_{\rm Jup}$) & $\gamma$ & LogUniform(1, 100) & $27_{-8}^{+12}$ \\
\qquad (i.e., the change point) &  &  &  \\
Mean of [Fe/H] distribution & $C_1$ & Uniform(-1, 1) & $0.17 \pm 0.02$ \\
\qquad for low-mass population  & & &  \\
Standard deviation of [Fe/H] distribution & $S_1$ & Uniform(0, 2)  & $0.12 \pm 0.01$ \\
\qquad for low-mass population  & & &  \\
Mean of [Fe/H] distribution & $C_2$ & Uniform(-1, 1) & $-0.03 \pm 0.05$ \\
\qquad for high-mass population  & & & \\
Standard deviation of [Fe/H] distribution & $S_2$ & Uniform(0, 2) & $0.10 \pm 0.04$ \\
\qquad for high-mass population  & & & 
\enddata
\tablecomments{Reported posteriors and their uncertainties are the median and $68 \%$ confidence interval of the resulting distribution.}
\end{deluxetable*}

{ The prior distributions for our hyperparameters are shown in Table~\ref{tab: results}. For a given system, indexed as $j$, we assume that ${\rm [Fe/H]}_t^j$ is given by the piecewise expression}
\begin{equation}
    {\rm [Fe/H]}_t^j \sim \begin{cases}
    \mathcal{N}(C_1, S_1) & \text{if } M_c^j \leq \gamma \\
    \mathcal{N}(C_2, S_2)  & \text{if } M_c^j > \gamma
    \end{cases},
\end{equation}
{ also referred to as a ``change-point'' model \citep{raftery1986bayesian, green1995reversible}. Here, $\mathcal{N}$ indicates a Normal distribution and $M_c$ is sampled directly from the posterior distribution generated by \citet{vanzandt2025cls}. The observed metallicity ${\rm [Fe/H]}_o^j$ is modeled as a random Gaussian variable with the form}
\begin{equation}
    {\rm [Fe/H]}_o^j \sim \mathcal{N} \left( {\rm [Fe/H]}_t^j, \sqrt{(\sigma_{{\rm [Fe/H]}_o^j})^2 + S^2} \right)
\end{equation}
{ such that the likelihood function for ${\rm [Fe/H]}_o^j$ is}
\begin{equation}
    \ln \mathcal{L}_{\rm [Fe/H]}^j = \frac{ \left( {\rm [Fe/H]}_t^j - \rm{[Fe/H]}_o^j \right)^2}{\sigma_{{\rm [Fe/H]}_o^j}^2 + S^2}
\end{equation}
{ where $S = S_1$ for $M_c^j \leq \gamma$ and $S = S_2$ for $M_c^j > \gamma$. The final likelihood of the model is evaluated as the sum of the natural logarithms across all systems:}
\begin{equation}
    \ln{\mathcal{L}} = \sum_{j=1}^N \ln{\mathcal{L}_{\rm [Fe/H]}^j}
\end{equation}
{ where $N$ is the number of systems in the sample.}

\begin{figure*}[t!]
  \centering
    \includegraphics[width=0.98\textwidth]{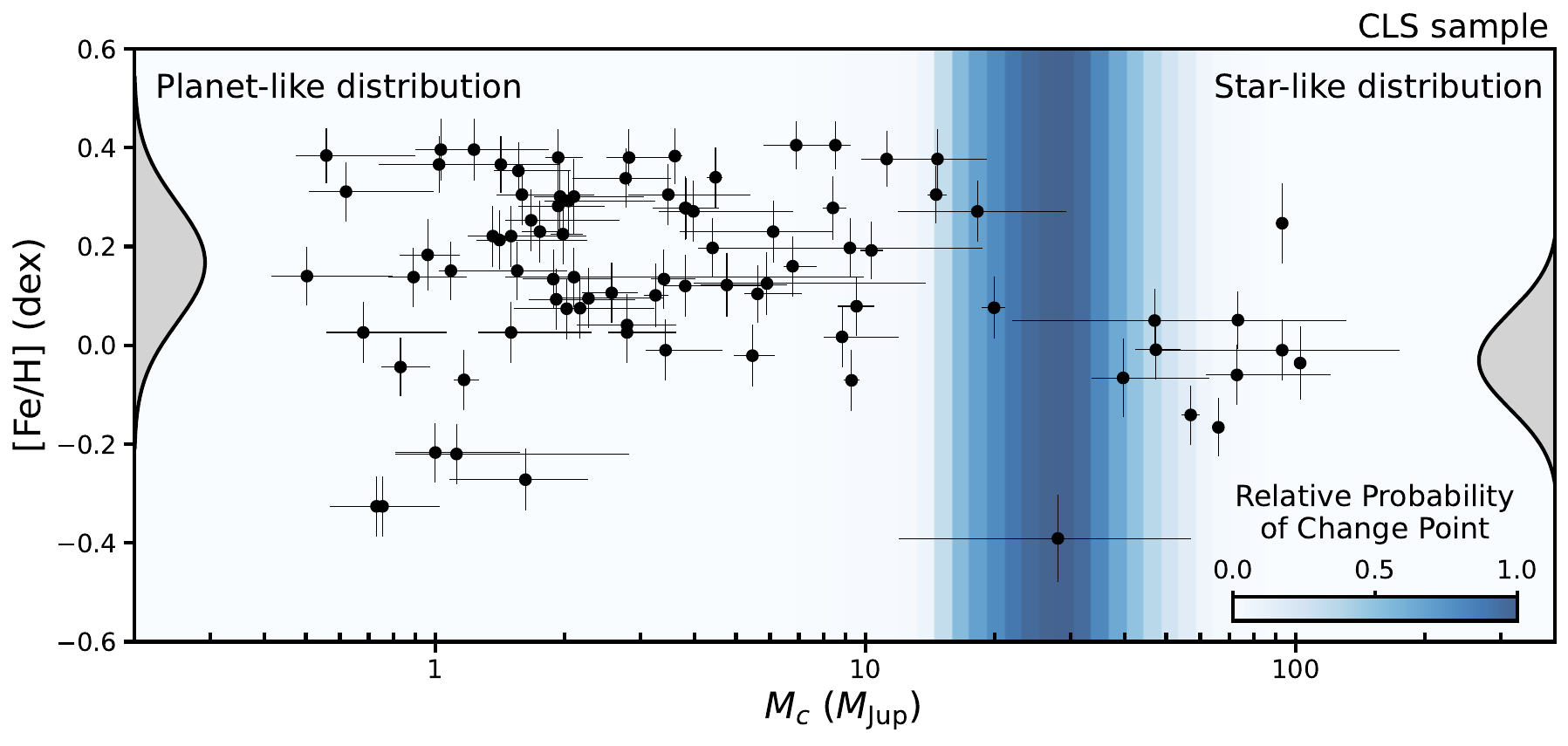}
    \includegraphics[width=0.98\textwidth]{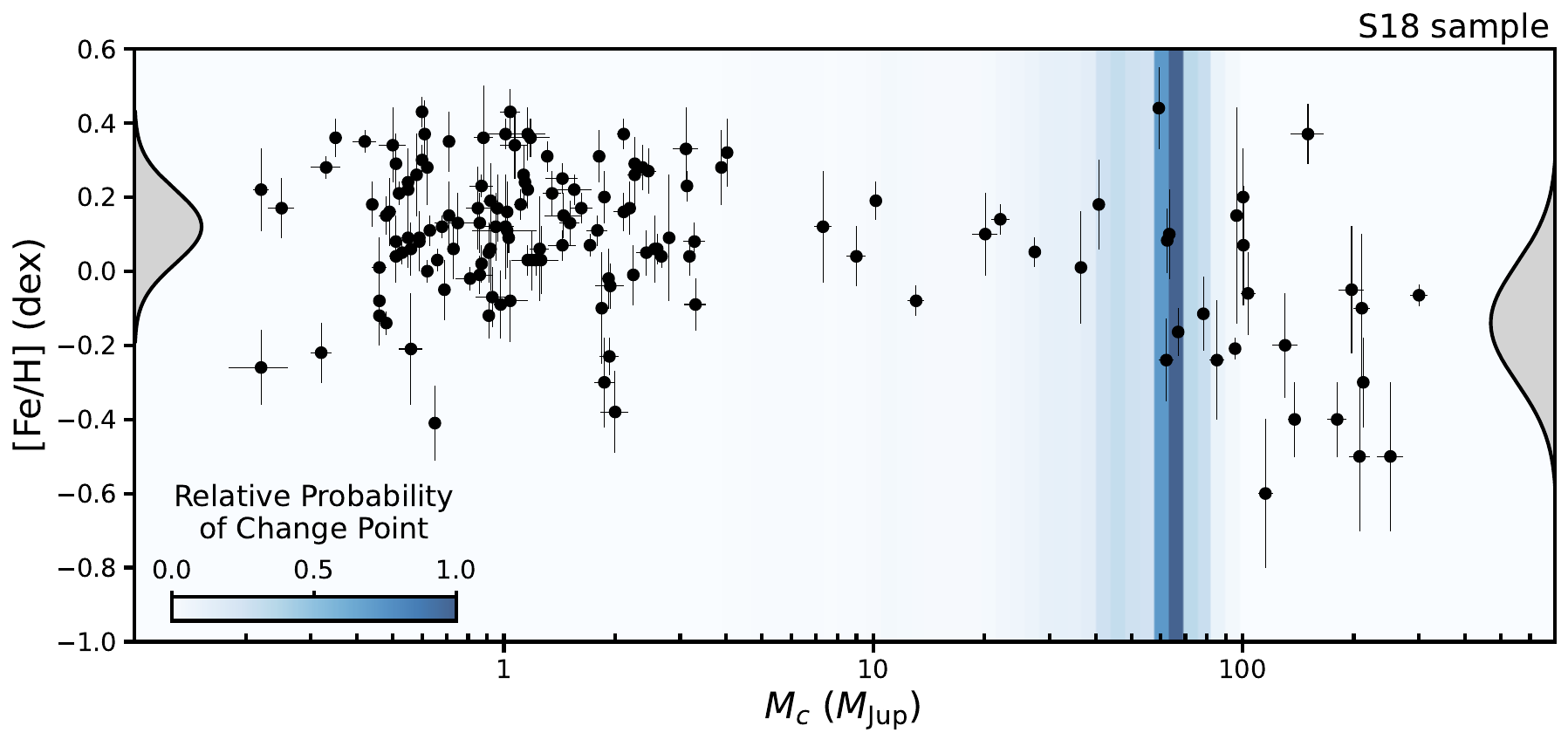}
    \caption{Stellar metallicity versus companion mass, displaying the results of the { HBM} described in Section \ref{sec: modelfit}. { The top panel shows the results of the model run on the CLS sample and the bottom panel shows the results of the model run on the \citetalias{schlaufman2018evidence} sample.} The vertical blue stripes are the aggregate posterior probability distribution of $\gamma$, where darker blues represent higher relative probabilities. { For the CLS sample,} we calculated a transition mass of $\gamma = 27_{-8}^{+12} \, M_{\rm Jup}$. { For the \citetalias{schlaufman2018evidence} sample, a relatively high transition mass of $\gamma = 59^{+10}_{-25} \, M_{\rm Jup}$} is favored. Gaussian distributions with the best-fit means and standard deviations of the low-mass and high-mass populations are shown in the margins. In short, { [Fe/H] distribution of the high-mass population is broad and closely resembles that of field stars, whereas the [Fe/H] distribution of the low-mass population is more tightly concentrated at higher values.}}
    \label{fig: 2}
\end{figure*}

{ We sampled our model via Markov-Chain Monte Carlo (MCMC) using PyMC \citep{abril2023pymc}. We ran the MCMC with 10 chains for $120,000$ steps each, discarding the first $20,000$ steps as burn-in. We ensured the convergence of the model using the Gelman-Rubin statistic ($\hat{r} < 1.1$; \citealt{gelman1992mcmc}.

This calculation yielded a transition mass of $\gamma = 27_{-8}^{+12} \, M_{\rm Jup}$ ($95 \%$ confidence interval of [15, 54] $M_{\rm Jup}$). According to the posterior distribution for $\gamma$, the probability that the transition occurs at a mass at or below $10 \, M_{\rm Jup}$ is $< 1\%$. We found the ${\rm [Fe/H]}$ distribution for the low-mass companions to have a mean of $C_1 = 0.17 \pm 0.02$ and a standard deviation of $S_1 = 0.12 \pm 0.01$ and the ${\rm [Fe/H]}$ distribution for the high-mass companions to have a mean of $C_2 = -0.03 \pm 0.05$ and a standard deviation of $S_2 = 0.10 \pm 0.04$. The latter is roughly consistent with the ${\rm [Fe/H]}$ distribution of field stars reported by \citet{rosenthal2021CLS}. Our results are displayed visually in the top panel of Figure~\ref{fig: 2}.}

\begin{figure*}[t!]
  \centering
    \includegraphics[width=0.98\textwidth]{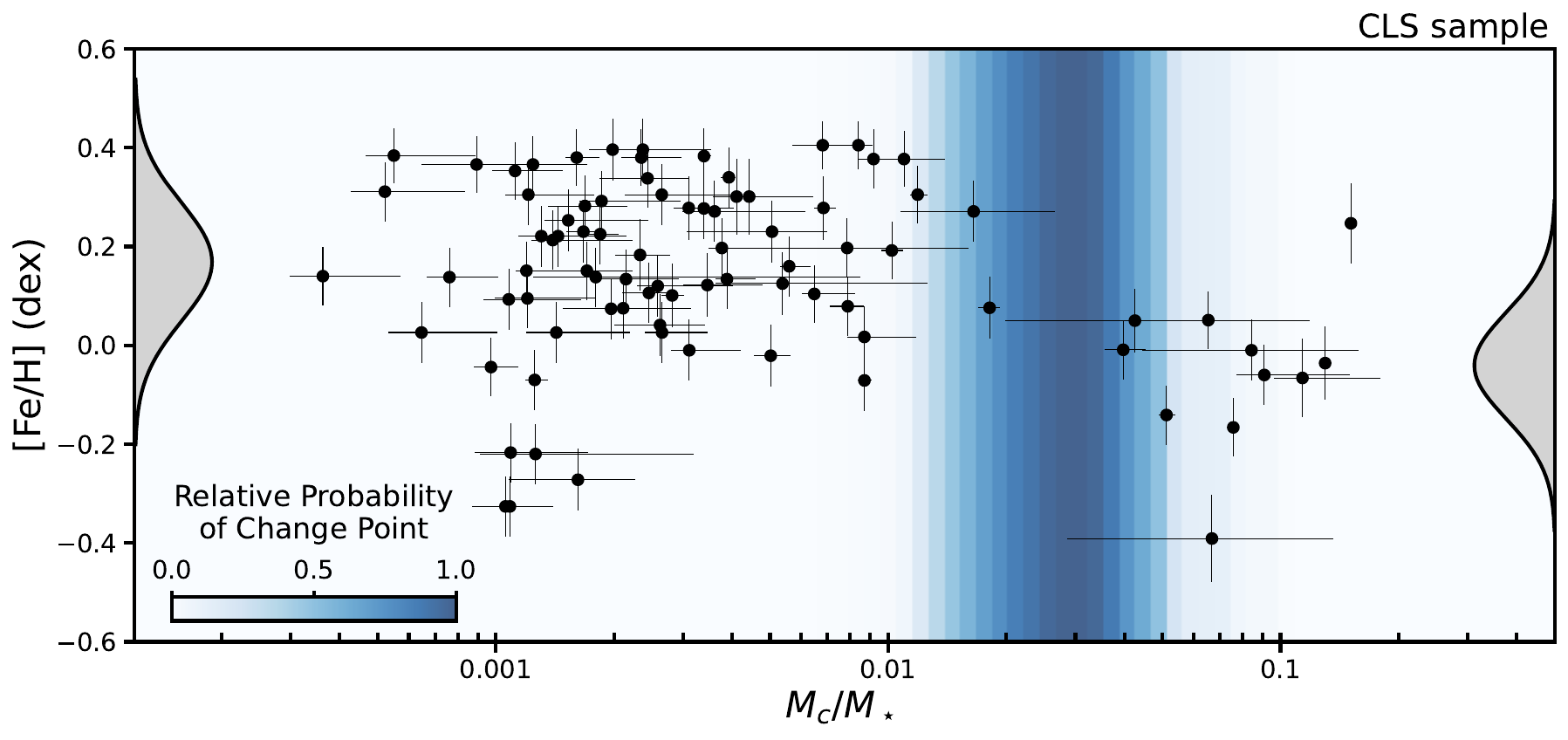}
    \caption{ Results of the HBM analysis on the CLS sample, but run in the plane of companion-host mass ratio rather than absolute companion mass (see Section~\ref{sec: massratio}). This analysis yielded a transition mass ratio of $0.027^{+0.015}_{-0.010}$.}
    \label{fig: massratio}
\end{figure*}

\subsection{ Transition as a Function of Mass Ratio}\label{sec: massratio}

{ We also run our model as a function of secondary-primary mass ratio, which may be more relevant than secondary mass alone in the context of planet formation. For this, we used the same HBM described above, but with $M_{c, {\rm post}}$ divided by the median value of the primary star mass (which has negligible uncertainty when compared to $M_c$). Our results for this analysis are shown in Figure~\ref{fig: massratio}. We recovered a transition mass ratio of $0.027^{+0.015}_{-0.010}$ ($95\%$ confidence interval of [0.013, 0.065]).}

\subsection{Tests for Statistical Significance}\label{sec: test}

To evaluate the significance of the change-point model, we compared it to a single-distribution model using the widely-applicable information criterion (WAIC; \citealt{watanabe2010waic}). The single-distribution model assumed that all of the data belongs to a single [Fe/H] distribution with some mean and standard deviation. The WAIC was calculated in PyMC using the computed log pointwise posterior predictive density, with a penalty applied to each model based on the number of parameters. We found that the WAIC for the change-point model exceeded that of the single-distribution model by $\gtrsim3$, meaning that the former was marginally favored over the latter.\footnote{ We came to the same conclusion using the leave-one-out cross validation, which is another model comparison method offered by PyMC.}

We also assessed the significance that the two samples are distinct using a two-sample Anderson-Darling test \citep{scholz1987ADtest}. We separated the data at an $M_c$ of $27 \, M_{\rm Jup}$, which is the median value of $\gamma$ determined by the HBM analysis. The two [Fe/H] distributions are shown in Figure~\ref{fig: 4}. This analysis yielded a two-sample Anderson-Darling test statistic of 8.96, which allowed us to reject the null hypothesis that the two samples come from the same distribution to a significance level $> 99.9 \%$.

Lastly, we repeated the exercise above using the two-sample Kolmogorov-Smirnov test \citep{hodges1958KStest}. This calculation yielded a p-value of $3.2 \times 10^{-5}$, again allowing us to reject the null hypothesis that the two samples come from the same distribution with high confidence.

\subsection{Evidence of Distinct Eccentricity Distributions}\label{sec: ecc}

\begin{figure*}[t!]
  \centering
    \includegraphics[width=0.98\textwidth]{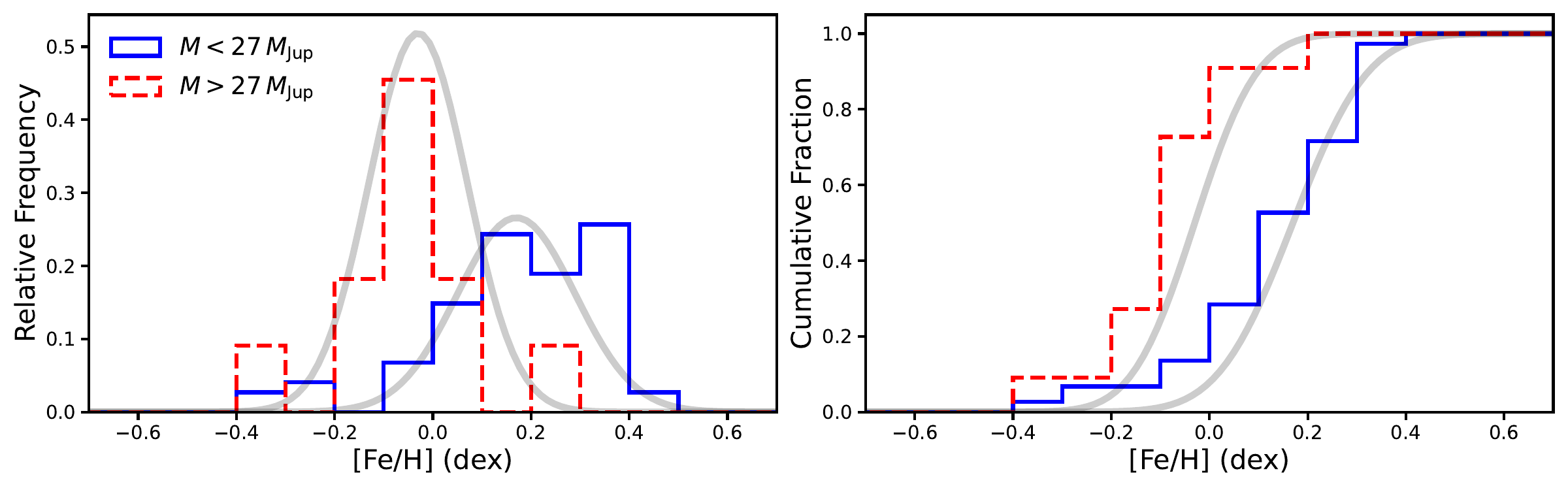}
    \includegraphics[width=0.98\textwidth]{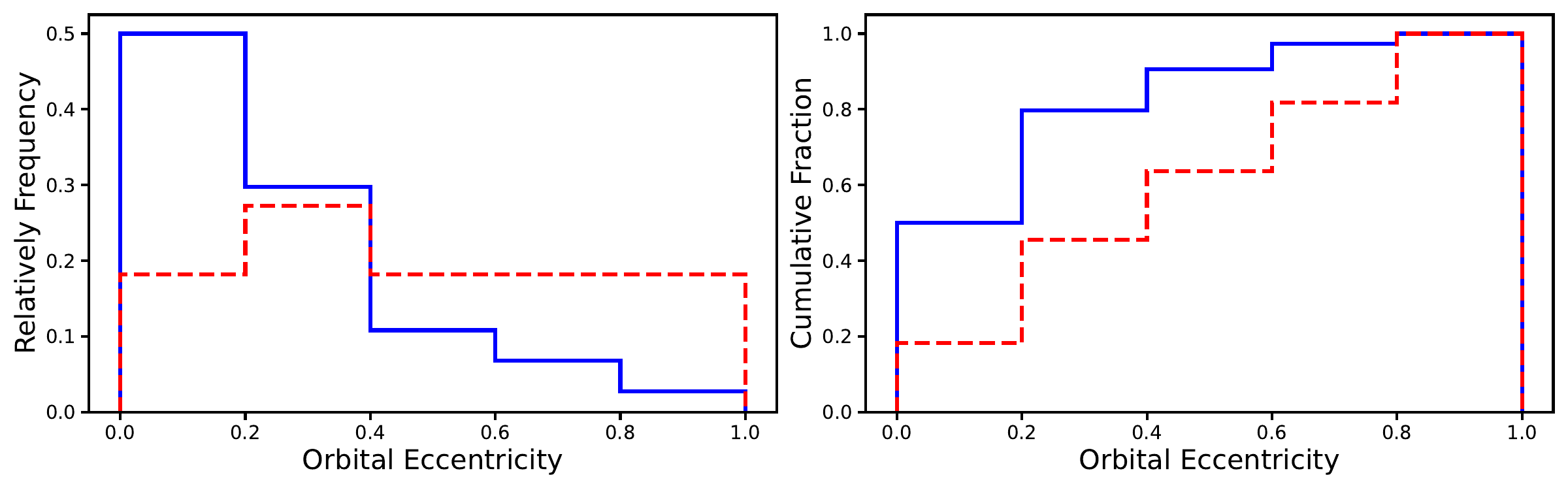}
    \caption{Histograms (left) and cumulative distributions (right) of stellar metallicity (top) and orbital eccentricity (bottom). The low-mass companion population is shown in blue and the high-mass companion population is shown in red. The gray curves are Gaussian distributions with the best-fit means and standard deviations calculated via the HBM analysis in Section \ref{sec: modelfit}. In Sections \ref{sec: test} and \ref{sec: ecc}, we find that the two populations have distinct stellar metallicity and orbital eccentricity distributions, according to two-sample Anderson-Darling and Kolmogorov-Smirnov tests.}
    \label{fig: 4}
\end{figure*}

Formation mechanism is also believed to influence orbital eccentricity; companions that form bottom-up are predicted to have relatively low orbital eccentricities compared to companions that form top-down \citep[e.g.,][]{veras2009formation, bate2012formation, duffell2015eccentric, grishin2015circularize}. We tested this prediction by comparing the eccentricity distributions on either end of our best-fit value of $\gamma$, which are shown in Figure \ref{fig: 4}. It was visually evident that the low-mass population has a higher probability of being in low-eccentricity orbits, whereas the high-mass population has a mean orbital eccentricity of $\sim 0.5$. A two-sample Anderson-Darling test of the two distributions (divided again at $M_c$ of $27 \, M_{\rm Jup}$) yielded a test statistic value of 4.05, allowing us to reject the null hypothesis that the two samples come from the same distribution to a significance level $> 99 \%$. A two-sample Kolmogorov-Smirnov test yielded a p-value of $0.03$, providing similar evidence. Thus, we found that the orbital eccentricity distributions of the two populations are indeed statistically distinct. { Nonetheless, we emphasize that just because the two eccentricity distributions are distinct when divided at 27 $M_{\rm Jup}$, it does not mean that this is the mass where the eccentricity distributions change. \citet{gilbert2025eccentricities} performed a more careful analysis of the eccentricities in the CLS sample and find that the eccentricity distribution changes gradually, rather than sharply, with increasing companion mass. We refer the reader to \citet{gilbert2025eccentricities} for more information on that analysis.}

% \newpage

\subsection{ Application to \citetalias{schlaufman2018evidence} Sample}

{ As a comparison, we ran a modified version of our HBM on the sample from \citetalias{schlaufman2018evidence}. The primary change to the model was the removal of the inclination term and the replacement of $M_c{\rm sin}(i)$ with mass, which was possible because all of the companions in \citetalias{schlaufman2018evidence} are transiting. This analysis yielded a transition mass of $\gamma = 59^{+10}_{-25} \, M_{\rm Jup}$, significantly higher than the value of $\sim 10 \, M_{\rm Jup}$ reported by \citetalias{schlaufman2018evidence}. For the [Fe/H] distributions, we found $C_1 = 0.12^{+0.02}_{-0.01}$, $S_1 = 0.10 \pm 0.01$, $C_2 = -0.12^{+0.06}_{-0.07}$, and $S_2 = 0.16^{+0.04}_{-0.03}$. These results are displayed in Figure~\ref{fig: 2}.}

\subsection{ Clustering Analysis}

{ As an independent check of our results, we repeated the three clustering analyses in \citetalias{schlaufman2018evidence}. Specifically, we utilized a hierarchical clustering algorithm (\texttt{scipy.cluster.hierarchy.linkage}), a k-means clustering algorithm (\texttt{scipy.cluster.vq.kmeans}), and Gaussian mixture clustering algorithm (\texttt{sklearn.mixture.GaussianMixture}). We performed each calculation $10^3$ times, each time resampling $M_c$ and [Fe/H] $10^3$ according to the observed values and their uncertainties. With each iteration, we recorded { both the most massive companion assigned to the low-mass population and the least massive companion assigned to the high-mass population. To determine the effective $1\sigma$ confidence interval of the transition mass, we calculated the $16^{\rm th}$ percentile of the former distribution and the $84^{\rm th}$ percentile of the latter distribution.}

The hierarchical clustering algorithm returned a transition mass range of 4 to 34 $M_{\rm Jup}$. The k-means clustering algorithm predicted a relatively low transition masses range of 2 to 11 $M_{\rm Jup}$. Lastly, the Gaussian mixture clustering algorithm predicted a transition mass range of 16 to 41 $\, M_{\rm Jup}$. { These values span a wide range and generally agree with the transition mass predicted by our HBM, with the exception of the k-means approach.}

\section{Discussion} \label{sec:discussion}

Our results suggest a transition in the formation mechanisms of substellar companions beyond 1~au at a mass of $27_{-8}^{+12} \, M_{\rm Jup}$. This estimate is higher than transition masses estimated by previous studies that analyzed stellar metallicities using different analysis techniques \citep{santos2017evidence, schlaufman2018evidence}, which reported transition masses of $4$ to $10 \, M_{\rm Jup}$. According to our $\gamma$ posterior distribution, the probability that the transition occurs at a mass under $10 \, M_{\rm Jup}$ is $< 1 \%$. We note that our estimate is broadly consistent with estimates of $\gamma$ obtained from other observational signatures. We outline these lines of evidence below.

\subsection{Occurrence Rates}

The earliest estimates of $\gamma$ were made based on the occurrence rates of giant planets and brown dwarfs. \citet{grether2006bd} analyzed the $M_c{\rm sin}(i)$ distribution of substellar companions detected by radial velocity surveys and found that it reaches a minimum at $M_c{\rm sin}(i) = 31^{+25}_{-18} \, M_{\rm Jup}$. \citet{sahlmann2011divide} later broke the orbital inclination degeneracy for this sample using Hipparcos astrometry and found that the minimum occurs between true masses of $25 \, M_{\rm Jup}$ and $45 \, M_{\rm Jup}$. { A similar minimum was found more recently by \citet{currie2023transition}.} These findings agree with our estimate of $\gamma$, suggesting that the transition in formation mechanism supported by our analysis of stellar metallicities may coincide with a local minimum in occurrence rate. { For an analysis of the occurrence rates of giant planets and brown dwarfs in the CLS sample, see \citet{vanzandt2025cls}.}

While free-floating giant planets and brown dwarfs likely experience different formations and evolution histories compared to those bound to stellar-mass primaries, appreciable effort has been spent to better understand their formation mechanisms \citep[e.g.,][]{luhman2012review, miretroif2023ffp, offner2023origin}. This effort has accelerated in recent years thanks to JWST \citep{pearson2023jumbo, langeveld2024jwst}. Recently, \citet{defurio2024turnover} found evidence of a turnover in the efficiency of free-floating brown dwarf formation in near-infrared observations of the young stellar cluster NGC 2024. From their analysis, the authors concluded that top-down formation (specifically those of core fragmentation) likely rapidly drops in efficiency below $13 \, M_{\rm Jup}$.

\subsection{Orbital Architectures}

The orbits of substellar companions also provide useful clues towards their formations. In particular, bottom-up formation is expected to produce relatively circular orbits compared to top-down formation due to interactions between the protoplanetary disk and the embedded companion \citep[e.g.,][]{veras2009formation, bate2012formation, duffell2015eccentric, grishin2015circularize}. In addition, bottom-up formation is predicted to result in orbits aligned with the stellar equator more frequently than top-down formation \citep[e.g.,][]{bate2010misalignment, offner2016misalignment}. A number of studies have investigated these predictions for eccentricity \citep{ma2014statistical, bowler2020eccentricity, vowell2025bd} and spin-orbit angle \citep[e.g.,][]{bryan2020obliquity, bryan2021obliquity, rice2022alignment, bowler2023obliquity, giacalone2024gpx1, poon2024obliquity, rusznak2024alignment, radzom2024alignment, smith2024alignment, wang2024alignment}. Estimates for $\gamma$ based on orbital eccentricity range from $< 10 \, M_{\rm Jup}$ to several tens of $M_{\rm Jup}$. Our somewhat simple analysis of orbital eccentricities finds support for a transition in eccentricities around ${\sim}27 \, M_{\rm Jup}$ (Figure \ref{fig: 4}), while a more careful analysis of the CLS eccentricity distribution in \citet{gilbert2025eccentricities} finds evidence for a gradual transition. We were refer the reader to that paper for a more in-depth discussion of this topic. Analyses of spin-orbit angles have found that close-in companions are more likely to be aligned with the equators of their host stars when higher in mass \citep{giacalone2024gpx1, radzom2024alignment, carmichael2025oatmealII, vowell2025oatmealIII}, whereas analyses of wide-separation companions have found the opposite \citep{bowler2023obliquity}. Evidently, the question of how to distinguish bottom-up from top-down formation on the basis of orbital architectures remains far from settled.

\subsection{Atmospheric Composition}

Perhaps the most powerful signature for distinguishing the formation mechanisms of substellar companions is atmospheric composition. The carbon-to-oxygen ratios and metallicities of substellar companion atmospheres are believed to be closely dependent on how and where they form \citep[e.g.,][]{oberg2011co, madhusudhan2012co, lothringer2021refrectory, chachan2023content}. In general, objects that form bottom-up should have atmospheres with higher carbon-to-oxygen ratios and metallicities than their host stars, whereas those that form top-down should have similar compositions to their host stars \citep{hawkins2020homogeneity}.

One of the earliest investigations of the boundary between bottom-up and top-down formation via atmospheric composition came from \citet{beatty2018overluminous}. Based on Spitzer secondary eclipse observations of the close-in, $36.5 \, M_{\rm Jup}$ companion to the star CWW~89A, they determined that the dayside of the companion is significantly more luminous than that predicted by cooling models. The authors argued that this overluminosity could be explained by a temperature inversion in the atmdosphere of the companion, which would the companion to have a carbon-to-oxygen higher than its host star. Because such high carbon-to-oxygen ratios are generally a result of bottom-up formation, the authors propose that the massive companion may have formed via the core-accretion mechanism. Atmospheric characterizations of other transiting companions with $M = 0.5 - 10 \, M_{\rm Jup}$ and $a < 0.1$~au have revealed a wide range of atmospheric carbon-to-oxygen ratios but often super-stellar metallicities \citep[e.g.,][]{alderson2023hj, august2023hj, bean2023jwst, bell2023jwst, brogi2023igrins, finnerty2024kpic, xue2024jwst}, generally supporting the notion that these close-in companions form bottom-up. Further investigation into the atmospheres of more massive transiting companions is needed to determine if this trend breaks at some critical mass.

Others have gained insight into the formations of more distant, directly imaged substellar companions by characterizing their atmospheres \citep[e.g.,][]{gravity2020co, miles2020bd, miles2023bd, wang2022bd, wang2023bd, inglis2024bd, phillips2024cloudy}. For companions beyond 10~au with masses between 50 and $70 \, M_{\rm Jup}$, measurements of atmospheric composition have generally been consistent with top-down formation \citep[e.g.,][]{line2015bd, wang2022bd, xuan2022bd, phillips2024bd}. Recently, \citet{hoch2023co} and \citet{xuan2024compositions} showed that the same is true for companions beyond 50~au with masses between 10 and $30 \, M_{\rm Jup}$. While atmospheric constraints are sparse for directly imaged companions with masses below 10~$M_{\rm Jup}$, early results suggest they are generally consistent with high metal enrichment and therefore bottom-up formation \citep[e.g.,][]{zhang2023elpis, zhang2025elpis}. { Notably, \citet{wang2025accretion} showed that directly imaged companions with masses below 10~$M_{\rm Jup}$ have metallicities higher than their host stars, whereas those with masses above 20~$M_{\rm Jup}$ have metallicities consistent with those of their host stars. Taken together, these studies suggest a transition mass in the range of $10$ to $20 \, M_{\rm Jup}$ at these very wide separations.} We will gain greater insight into the formations of giant planet and brown dwarfs in the $1 - 10$~au range as direct imaging techniques improve, allowing us to resolve substellar companions at smaller angular separations.

As of today, few atmospheric investigations of giant planets and brown dwarfs near the water-ice line have been conducted. Only a small number of known giant planets transit at these orbital separations, making transmission spectroscopy infeasible for most systems \citep[e.g.,][]{alam2022hip41378}. \textit{Gaia} is predicted to increase this number by several dozen, enabling investigation of temperate substellar companions at a statistical level \citep{perryman2014gaia}. Direct imaging provides another promising route to understanding the origins of these objects, as surveys have gradually improved sensitivity to low-mass companions in the $1-10$~au range around young stars \citep[e.g.,][]{franson2023aflep, hinkley2023hd206893}. In the coming years, space-based coronagraphy with JWST, Roman, and planned future missions will begin to shed light on their atmospheric contents \citep{franson2024aflep}.

% \newpage

\subsection{ Implications for Planet Formation Theory}

{ Most planet formation models struggle to produce giant planets with masses of $\sim 20 \, M_{\rm Jup}$ in typical protoplanetary disks \citep{tanigawa2007finalmass, mordasini2009a, mordasini2009b, rosenthal2020finalmass, adams2021formation}. In the simplest terms, the maximum mass attainable by a planet can be estimated as a function of the disk viscosity $\alpha$ parameter \citep{shakura1973alpha}, the initial disk mass, and the disk depletion timescale \citep{tanigawa2016formation}. \citetalias{schlaufman2018evidence} thoroughly explored these parameters and demonstrated that very massive planets form more easily when $\alpha$ is lower, initial disk mass is higher, or disk depletion timescale is longer.

Observational constraints on these parameters allow us to assess the maximum planetary mass heuristically. Estimates of $\alpha$ span many orders of magnitude \citep[e.g.,][]{rafikov2017viscous, fedele2018almagaps}, with values ranging from 0.1 to $\leq 10^{-4}$. Disk depletion timescales are well constrained to $< 10$~Myr for most Sun-like stars \citep{strom1989disk, haisch2001disk, williams2011disklifetime}. Measurements of initial disk mass, however, are more challenging. For instance, \citet{ansdell2016} performed an ALMA survey of protoplanetary disks around primarily low-mass ($0.1 - 0.5 \, M_\odot$) stars in the $1-3$~Myr-old Lupus complex and found most disks to be less massive than Jupiter, concluding that most of the gas and dust was depleted early in the disk lifetime due to planet formation. 

If we assume low values of $\alpha$ and typical disk depletion rates, high initial disk masses are required to form the most massive planets. \citet{tanaka2020finalmass} predicted that planets with masses up to $\sim 20 \, M_{\rm Jup}$ can form in disks with $\alpha = 10^{-3}$ and initial masses of $\sim 0.1 \, M_\star$ (roughly $10 \times$ higher than the minimum mass solar nebula, assuming a Sun-like host star). { There is tentative evidence that some disks can reach these initial masses \citep{speedie2024disk, speedie2025disk}}, and the fact that companions with masses of $10 - 20 \, M_{\rm Jup}$ are so rare means that high initial masses are only required in a small fraction of disks to reproduce the observed planetary mass function (see \citealt{vanzandt2025cls} for more info). In other words, a maximum planet mass of $\sim 27 \, M_{\rm Jup}$ is consistent with the notion that $10 - 20 \, M_{\rm Jup}$ planets represent the tail end of planet formation, where formation efficiency is low due to the small fraction of disks with the appropriate initial conditions.}

\section{Conclusions} \label{sec:conclusion}

Using a hierarchical Bayesian model, we inspected the distribution of giant planets, brown dwarfs, and low-mass stellar companions with orbital separations beyond 1~au detected by the California Legacy Survey \citep{rosenthal2021CLS} in the companion mass -- stellar metallicity plane. We found that the sample is well-described by a change-point model, in which two distinct stellar metallicity distributions are separated by a transition mass, $\gamma$. We calculated the transition mass to be $\gamma = 27_{-8}^{+12} \, M_{\rm Jup}$, with relatively low-mass companions preferring to orbit relatively metal-rich stars ($\rm{[Fe/H]} = 0.17 \pm 0.12$ dex) and relatively high-mass companions orbiting relatively metal-poor stars ($\rm{[Fe/H]} = -0.03 \pm 0.10$ dex). In addition, we found that the low-mass population has a stronger preference for low-eccentricity orbits (see \citealt{gilbert2025eccentricities} for a more careful analysis of eccentricities). Overall, these findings are consistent with predictions of substellar companion formation: those with relatively low masses form bottom up, whereas those with relatively high-masses for top down. Our estimate of $\gamma$ broadly agrees with previous estimates based on other metrics.

% Our estimate of $\gamma$ broadly agrees with previous estimates based on other metrics, but hints that $\gamma$ may be larger near the water-ice line than it is for companions on relatively close-in orbits. According to our analysis, there is a $90 \%$ probability that our sample has an $\gamma$ higher than $10 \, M_{\rm Jup}$, the transition mass reported by \cite{schlaufman2018evidence} for companions within 1~au. If this is correct, it may mean that bottom-up formation can produce more massive planets near the water-ice line, where the incidence of giant planets also peaks \citep[e.g.,][]{cumming2008distribution, bowler2010distribution, bryan2016statistics, fernandes2019turnover, fulton2021CLS}. Alternatively, it may suggest that top-down formation is inefficient at these orbital separations, leading to a dearth of low-mass brown dwarfs. 

Our analysis utilized a large sample of important, homogeneously characterized stars in the Solar neighborhood. The results presented herein will have important implications for our understanding of discoveries made by current and future missions capable of probing faint companions near the water-ice line. For instance, \textit{Gaia} DR4 is predicted to contain thousands of giant planets and brown dwarfs between 1 and 5~au \citep{perryman2014gaia}. Our estimate of $\gamma$ will serve as a useful prior as we begin to uncover trends in demographics and orbital properties for these systems. Eventually, the Habitable Worlds Observatory and Large Interferometer For Exoplanets will probe Sun-like stars in the Solar neighborhood for planets, enabling detailed calculations of atmospheric compositions via reflected and emitted light \citep[e.g.,][]{quanz2022LIFE, carriongonzalez2023LIFE, alei2024LIFE}. Our results place informative constraints on the physics of planet formation near the habitable zone, where these missions will search for and characterize terrestrial planets.

\begin{acknowledgements}

{ We thank the anonymous referee for their comments that significantly strengthened this paper.} We also thank Heather Knutson, Howard Isaacson, and Benjamin J. Fulton for their constructive conversations and feedback.

SG is supported by an NSF Astronomy and Astrophysics Postdoctoral Fellowship under award AST-2303922.

\end{acknowledgements}

\software{\texttt{PyMC} \citep{abril2023pymc}, \texttt{SciPy} \citep{virtanen2020SciPy}, \texttt{scikit-learn} \citep{scikit-learn}}
 
\bibliography{bibliography}{}
\bibliographystyle{aasjournal}

\end{document}